\documentclass[showpacs, secnumarabic, amssymb, nobibnotes, aps, prb, reprint]{revtex4-1}
\usepackage{graphicx}%
\usepackage{dcolumn}%
\usepackage{bm}%
\usepackage{lineno}

\begin{document}

\title{Spinodal Decomposition of Magnetic Ions in Eu-Codoped Ge$_{1\textrm{-}x}$Cr$_{x}$Te}

\author{A.~Podg\'{o}rni}
\email[Electronic mail: ]{podgorni@ifpan.edu.pl}
\author{L.~Kilanski}
\author{W.~Dobrowolski}
\author{M.~G\'{o}rska}
\author{A.~Reszka}
\author{V.~Domukhovski}
\author{B.~J.~Kowalski}
\author{B.~Brodowska}
\affiliation{Institute of Physics, Polish Academy of Sciences, al. Lotnikow 32/46, 02-668 Warsaw, Poland}

\author{J.~R.~Anderson}
\author{N.~P.~Butch}
\altaffiliation[Present address: ]{Lawrence Livermore National Laboratory, 7000 East Avenue, Livermore, CA 94550, USA}
\affiliation{Department of Physics and Center for Nanophysics and Advanced Materials, University of Maryland, College Park, MD 20742, USA}

\author{V.~E.~Slynko}
\author{E.~I.~Slynko}
\affiliation{Institute of Materials Science Problems, Ukrainian Academy of Sciences, 5 Wilde Street, 274001 Chernovtsy, Ukraine}

\date{\today}

\begin{abstract}

We present the experimental evidence for the presence of spinodal decomposition of the magnetic ions in the Ge$_{1\textrm{-}x\textrm{-}y}$Cr$_{x}$Eu$_{y}$Te samples with chemical composition varying in the range of 0.015$\,$$\leq$$\,$$x$$\,$$\leq$$\,$0.057 and 0.003$\,$$\leq$$\,$$y$$\,$$\leq$$\,$0.042. The ferromagnetic transition at temperatures 50$\,$$\leq$$\,$$T$$\,$$\leq$$\,$57$\;$K was observed, independent of the chemical composition. The long-range carrier mediated itinerant magnetic interactions seem to be responsible for the observed ferromagnetic order. The magnetic irreversibility with coercive field $H_{C}$$\,$$=$$\,$5$...$63$\;$mT and the saturation magnetization $M_{S}$$\,$$\approx$$\,$2$...$6$\;$emu/g are found to strongly depend on the chemical composition of the alloy.

\end{abstract}

\keywords{semimagnetic-semiconductors; ferromagnetic-materials;
magnetic-properties}

\pacs{72.80.Ga, 75.40.Cx, 75.40.Mg, 75.50.Pp}

% 72.80.Ga Transition-metal compounds
% 75.40.Cx Static properties order parameter, static susceptibility, heat capacities, critical exponents, etc.
% 75.50.Pp Magnetic semiconductors
% 75.40.Mg Numerical simulation studies

%\end{frontmatter}

\maketitle

\linenumbers

\begin{figure}
  \begin{center}
    \includegraphics[width = 0.5\textwidth, bb = 70 110 780 510]
    {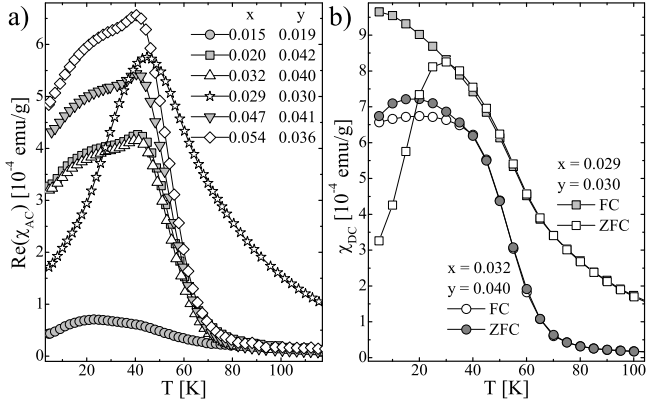}\\
  \end{center}
  \caption{\small (a) The real part of the ac magnetic susceptibility and (b) the dc magnetic susceptibility as a function of temperature for the Ge$_{1\textrm{-}x\textrm{-}y}$Cr$_{x}$Eu$_{y}$Te samples with different chemical composition.}
  \label{FigMagnProp}
\end{figure}

\begin{figure}[h!]
  \begin{center}
    \includegraphics[width = 0.42\textwidth, bb = -10 40 610 560]
    {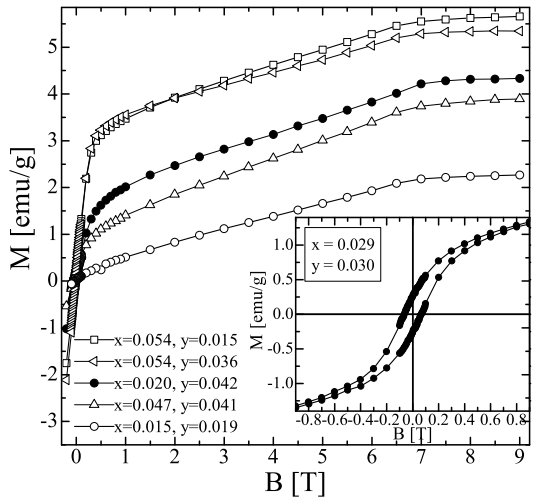}\\
  \end{center}
  \caption{\small The isothermal magnetization curves observed at $T$$\,$$=$$\,$4.5$\;$K for the selected Ge$_{1\textrm{-}x\textrm{-}y}$Cr$_{x}$Eu$_{y}$Te samples with different chemical composition. The inset shows the exemplary magnetic hysteresis curve obtained at $T$$\,$$=$$\,$4.5$\;$K for the selected crystal.}
  \label{FigMHCurves}
\end{figure}

\begin{figure}[h!]
  \begin{center}
    \includegraphics[width = 0.5\textwidth, bb = 0 20 660 520]
    {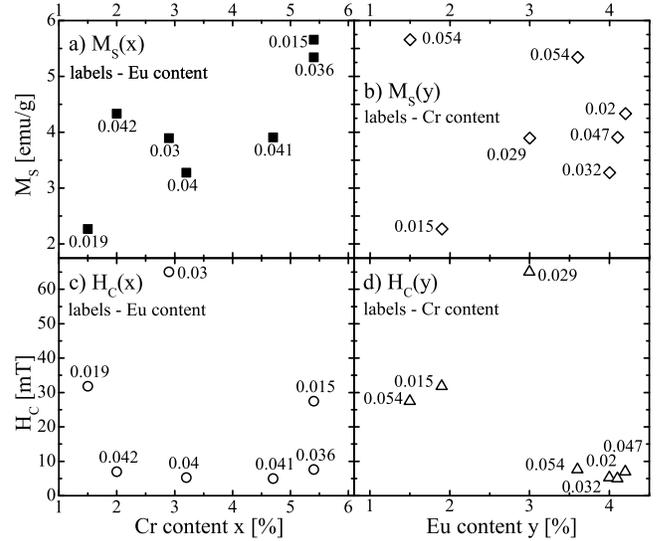}\\
  \end{center}
  \caption{\small The saturation magnetization $M_{S}$ and coercive field $H_{C}$ as a function of the amount of magnetic ions $x$ and $y$ obtained at $T$$\,$$=$$\,$4.5$\;$K for Ge$_{1\textrm{-}x\textrm{-}y}$Cr$_{x}$Eu$_{y}$Te samples with different chemical composition.}
  \label{FigMSHCvsXY}
\end{figure}

Semiconductor spintronics remains still a technological challenge requiring development of ferromagnetic semiconductors with the Curie temperature above 300 K.\cite{Ohno10a} In the transition metal doped IV-VI materials the itinerant ferromagnetism with relatively high Curie temperatures is easily achievable because of the high Mn-hole magnetic exchange constant, $J_{pd}$$\,$$\approx$$\,$0.8$\;$eV, (Ref.$\;$\onlinecite{Cochrane74a}) and natural high $p$-type conductivity with carrier concentration,  $n$$\,$$\approx$$\,$10$^{20}$$...$10$^{21}$$\;$cm$^{-3}$, (Ref.$\;$\onlinecite{Kilanski09a}). The highest reported Curie temperature among IV-VI materials is around 200$\;$K for Ge$_{1\textrm-x}$Mn$_{x}$Te with $x$$\,$$=$$\,$0.5 (Ref.$\;$\onlinecite{Lechner10a}) and 160$\;$K for Ge$_{1\textrm-x}$Cr$_{x}$Te for $x$$\,$$\approx$$\,$0.06 (Ref.$\;$\onlinecite{Fukuma07a}), close and slightly above the values for Ga$_{1\textrm-x}$Mn$_{x}$As and much higher than other III-V and II-VI diluted magnetic semiconductors.\cite{Dietl10a} Thus, GeTe based diluted magnetic semiconductors are perceived as potential material for the semiconductor spintronics working at room temperature. In contrast to thin layers, bulk Ge$_{1\textrm-x}$Cr$_{x}$Te crystals show much lower transition temperatures, $T$$\,$$\leq$$\,$60$\;$K, with the presence of both ferromagnetic and spin-glass-like states.\cite{Kilanski11a,Kilanski12a} The nature of this decrease of the magnetic ordering temperature remains an unexplained issue. \\
\indent The present paper undertakes the problem of spinodal decomposition of magnetic ions influencing greatly the magnetic properties of the Ge$_{1\textrm{-}x\textrm{-}y}$Cr$_{x}$Eu$_{y}$Te alloy. Here we continue our previous investigations devoted to Ge$_{1\textrm{-}x}$Cr$_{x}$Te alloy showing spin-glass and ferromagnetic order at $T$$\,$$<$$\,$60$\;$K (see Refs.$\;$\onlinecite{Kilanski11a,Kilanski12a}). In the present work we show, that irrespective of the average Cr content, the Curie temperature can be stabilized by the average Eu content. The other parameters describing the magnetic properties of a magnetic material, such as coercive field, $H_{C}$, and saturation magnetization, $M_{S}$, show changes with the average Cr content. \\
\indent We present the studies of Ge$_{1\textrm{-}x\textrm{-}y}$Cr$_{x}$Eu$_{y}$Te crystals grown by the modified Bridgman method. The chemical composition of our samples was determined by using the energy dispersive x-ray fluorescence method (EDXRF). The relative error of the chemical composition determination was not exceeding 10$\%$. The EDXRF measurements showed that the chemical content of the Ge$_{1\textrm{-}x\textrm{-}y}$Cr$_{x}$Eu$_{y}$Te samples changed in the range of 0.015$\,$$\leq$$\,$$x$$\,$$\leq$$\,$0.057 and 0.003$\,$$\leq$$\,$$y$$\,$$\leq$$\,$0.042. \\
\indent The X-ray diffraction measurements were done with the use of Siemens D5000 diffractometer at room temperature. The experimental diffraction patterns were fitted with the use of Rietveld refinement method in order to determine the crystallographic parameters of the samples. The obtained results indicated, that our samples were single phased and crystallized in the rhombohedrally distorted GeTe structure with the lattice parameter $a$$\,$$\approx$$\,$5.98$\;$$\textrm{\AA}$ and the angle of distortion $\alpha$$\,$$\approx$$\,$88.3$\;$deg, similarly to the parameters for pure GeTe (see Ref.$\;$\onlinecite{Galazka99a}).
\begin{figure}
  \begin{center}
    \includegraphics[width = 0.5\textwidth, bb = 50 50 810 580]
    {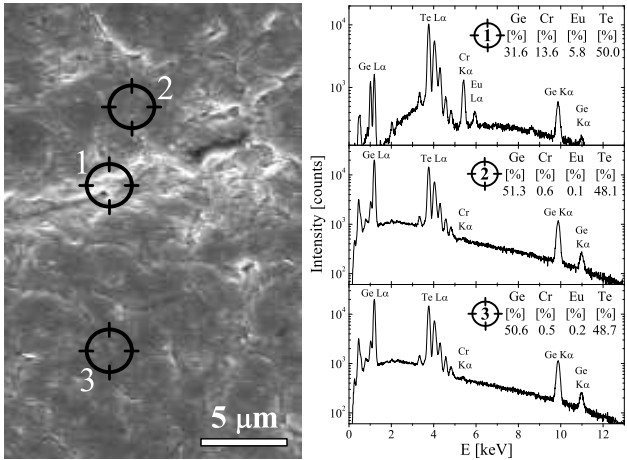}\\
  \end{center}
  \caption{\small (a) The SEM image of the Ge$_{0.912}$Cr$_{0.047}$Eu$_{0.041}$Te sample and (b) EDX spectra measured at the selected spots of the samples (marked with crossbars).}
  \label{FigSEMEDS}
\end{figure}
\\
\indent A series of high-resolution micrographs was collected for the studied Ge$_{1\textrm{-}x\textrm{-}y}$Cr$_{x}$Eu$_{y}$Te samples with the use of Hitachi UHR-FE-SEM (SEM) scanning electron microscope coupled with the energy dispersive X-ray spectrometry (EDX). Our studies showed the presence of spinodal decomposition of Cr ions in all the studied samples. The exemplary image of the sample surface together with the EDX spectra gathered at selected sample spots are presented in Fig.$\;$\ref{FigSEMEDS}. The data presented in Fig.$\;$\ref{FigSEMEDS} are typical of our samples, similar features were observed in several crystals with different chemical composition. The SEM maps showed strong chemical contrast indicating the presence of regions with different chemical content than an average EDXRF data. The EDX spectra gathered from selected spots of Fig.$\;$\ref{FigSEMEDS} indicated the presence of spinodal-decompositions into regions with low (point No. 1) and high (points No. 2 and 3) transition-metal content. The average size of the chemical inhomogeneities present in our samples changed from 1$\;$$\mu$m up to 10$\;$$\mu$m. The average chemical composition observed for the Ge$_{0.912}$Cr$_{0.047}$Eu$_{0.041}$Te sample (see Fig.$\;$\ref{FigSEMEDS}) was not exceeding $x$$\,$$\approx$$\,$0.006 and $y$$\,$$\approx$$\,$0.002. The maximum dilution limit of transition-metal ions in our samples can be estimated to not exceed $x$$\,$$\approx$$\,$0.012 and $y$$\,$$\approx$$\,$0.005. The presence of Cr-related nanoclusters is known in literature to be responsible for room temperature ferromagnetism in Zn$_{1-x}$Cr$_{x}$Te diluted magnetic semiconductors.\cite{Kuroda07a} However, the nature of nanoclusters present in our samples seems to be different than the presence of CrTe related phases, such as Cr$_{7}$Te$_{8}$, Cr$_{5}$Te$_{6}$, and Cr$_{3}$Te$_{4}$ with $T_{C}$$\,$$>$$\,$300$\;$K (Ref.$\;$\onlinecite{Kuroda07a}), since we do not observe room temperature ferromagnetism. The magnetic properties of our samples are also different than the ones reported for Cr$_{5}$Te$_{8}$ (with 190$\,$$>$$\,$$T_{C}$$\,$$>$$\,$245$\;$K, Ref.$\;$\onlinecite{Lukoschus04a}), Cr$_{2}$Te$_{3}$ (with $T_{C}$$\,$$=$$\,$170$\;$K, Ref.$\;$\onlinecite{Hashimoto71a}), and CrTe$_{2}$ (with $T_{C}$$\,$$=$$\,$18$\;$K, Ref.$\;$\onlinecite{Zhang90a}). It is more probable, that the large size of our inhomogeneities and high carrier concentration, $n$$\,$$>$$\,$10$^{20}$$\;$cm$^{-3}$, can cause the long range itinerant carrier-mediated magnetic interactions to be the most significant in our samples. \\
\indent The magnetotransport studies including the resistivity and the Hall effect measurements were performed at room temperature using the standard six-contact dc-current Hall configuration and constant magnetic field $B$$\,$$=$$\,$1.5$\;$T. The results of magnetotransport studies revealed that all the studied Ge$_{1\textrm{-}x\textrm{-}y}$Cr$_{x}$Eu$_{y}$Te samples had typical features for GeTe-based materials,\cite{Kilanski09a} i.e. $p$-type conductivity with relatively high carrier concentration, $p$$\,$$>$$\,$10$^{20}$$\;$cm$^{-3}$, and low mobility, $\mu$$\,$$<$$\,$100$\;$cm$^{2}$/(V$\cdot$s). The magnetotransport data were chemical composition independent, indicating that the magnetic impurities were not the main source of electrically active defects in our samples and the intrinsic cation vacancy concentration was higher than that possibly induced by the presence of magnetic ions. \\
\indent The magnetic properties of our Ge$_{1\textrm{-}x\textrm{-}y}$Cr$_{x}$Eu$_{y}$Te samples were studied with the use of a LakeShore 7229 susceptometer/magnetometer system. At first the measurements of the ac magnetic susceptibility were performed at temperatures 4.5$\,$$\leq$$\,$$T$$\,$$\leq$$\,$200$\;$K. During the measurement the sample was placed in the alternating magnetic field with amplitude $H_{AC}$$\,$$=$$\,$1$\;$mT and frequency $f$$\,$$=$$\,$625$\;$Hz. The temperature dependence of the real part of the magnetic susceptibility, Re($\chi$)($T$), for a few selected samples is presented in Fig.$\;$\ref{FigMagnProp}a. We observe a well defined increase of the magnetic susceptibility with decreasing temperature between 60 and 50$\;$K with a maximum around 50$\;$K, indicating a presence of a magnetic transition in the material. The frequency dependent measurements of the ac magnetic susceptibility showed a shift of the maximum in the Re($\chi_{AC}$) curves towards higher temperatures with increasing $f$ only in one sample with $x$$\,$$=$$\,$0.029 and $y$$\,$$=$$\,$0.030. Thus, we identify the observed ac susceptibility data as paramagnet-ferromagnet phase transition for all the samples except the Ge$_{0.941}$Cr$_{0.029}$Eu$_{0.030}$Te crystal, in which the mictomagnetic state is found. Possibly the magnetic order observed in our samples is due to the carrier mediated magnetic interactions. \\
\indent The dc magnetization measurements were performed over the temperature range 5$\,$$\leq$$\,$$T$$\,$$\leq$$\,$300$\;$K in magnetic fields $B$$\,$$=$2, 5, 10 and 20$\;$mT with the use of a Quantum Design superconducting quantum interference device (SQUID) MPMS XL-7 magnetometer. The linear behavior of the $M$($B$) curves within the studied field values allowed the calculation of the temperature dependencies of both zero-field-cooled (ZFC) and field-cooled (FC) dc magnetic susceptibility $\chi_{DC}$($T$). The obtained $\chi_{DC}$($T$) curves for the selected Ge$_{1\textrm{-}x\textrm{-}y}$Cr$_{x}$Eu$_{y}$Te samples are gathered in Fig.$\;$\ref{FigMagnProp}b. The bifurcations between the ZFC and FC $\chi_{DC}$($T$) curves were observed in all samples. It is a clear signature that the large frustration of the magnetic ions occurred in the studied system.
The isothermal high field magnetization curves $M$($B$) were studied over the temperature range 4.5$\,$$\leq$$\,$$T$$\,$$\leq$$\,$100$\;$ and up to a magnetic field $B$$\,$$=$$\,$9$\;$T with the use of Weiss extraction method employed into the LakeShore 7229 magnetometer system. The obtained magnetization curves at temperature $T$$\,$$=$$\,$4.5$\;$K for a few selected Ge$_{1\textrm{-}x\textrm{-}y}$Cr$_{x}$Eu$_{y}$Te samples are presented in Fig.$\;$\ref{FigMHCurves}.
The results of the measurements show that the magnetization was almost a linear function of the magnetic field for 1$\,$$\leq$$\,$$B$$\,$$\leq$$\,$8$\;$T. It is a direct signature of the strong magnetic frustration in the studied system. The saturation magnetization, $M_{S}$ , was reached for all our Ge$_{1\textrm{-}x\textrm{-}y}$Cr$_{x}$Eu$_{y}$Te samples at the highest used magnetic field values $B$$\,$$\geq$$\,$8$\;$T. The saturation magnetization value (see Figs.$\;$\ref{FigMSHCvsXY}a and $\;$\ref{FigMSHCvsXY}b) is an increasing function of the Cr content $x$, except for $x$$\,$$=$$\,$0.020 and $x$$\,$$=$$\,$0.029, while taking into account the Eu content $y$ there is no well defined correlation between $y$ and $M_{S}$. It is a signature that the Cr-alloying introduced more magnetically active ions than Eu, i.e. larger fraction of the Cr ions was substituting the cation sublattice sites in the host GeTe lattice. This is most apparent for the samples with $x$$\,$$\approx$$\,$0.05 and 0.015$\,$$<$$\,$$y$$\,$$<$$\,$0.041, where the saturation magnetization decreases with increasing amount of Eu in the alloy. It is very likely that with increasing amount of Eu it is distributed in the sites of the crystal lattice that promote the antiferromagnetic interactions between the paramagnetic ions. This may lead to the changes of the charge state of both Cr and Eu which then result in reduction of their spin state.
In this way the significant reduction of saturation magnetization of the material even in the growing amount of paramagnetic impurities in the alloy might be observed. \\
\indent The magnetization hysteresis loops were measured for all our samples at temperatures lower than $T$$\,$$<$$\,$100$\;$K. An appearance of well defined magnetic hysteresis loops is observed for all the samples (exemplary result shown in the inset to Fig.$\;$\ref{FigMHCurves}) at temperatures lower than the critical temperatures determined from the ac-susceptibility results. The observed shape of the hysteresis loops might originate from some changes of the geometrical  parameters of the spinodal decompositions present in our samples. The analysis of hysteresis loops showed that there exists a correlation between the coercive field $H_{C}$ values obtained at $T$$\,$$=$$\,$4.5$\;$K and the chemical composition of the samples (see Figs.$\;$\ref{FigMSHCvsXY}c and $\;$\ref{FigMSHCvsXY}d). The coercive field changed in the range of 5$\,$$<$$\,$$H_{C}$$\,$$<$$\,$63$\;$mT for our samples. The $H_{C}$ changed nonmonotonically with increasing Cr content, $x$, and was roughly a decreasing function of Eu content, $y$, except for $y$$\,$$=$$\,$0.03. It indicated, that poor allocation of Eu ions changed significantly the domain structure of the material allowing the control of its magnetic hardness for more than an order of magnitude. \\
\indent The presence of magnetic-ion-rich regions was observed in the case of the Ge$_{1\textrm{-}x\textrm{-}y}$Cr$_{x}$Eu$_{y}$Te samples with different chemical composition. The magnetic properties show that the transition temperatures in our samples are stabilized by the presence of magnetic Ge$_{1\textrm{-}x\textrm{-}y}$Cr$_{x}$Eu$_{y}$Te spinodal decompositions. The magnetic hardness of the alloy seems to be somewhat controlled by the amount of Eu in the samples leading to a decrease of $H_{C}$ from 63$\;$mT down to about 5$\;$mT for $y$$\,$$=$$\,$0.015 and 0.042, respectively.

\begin{acknowledgments}
\noindent The research was supported by the Foundation for Polish Science - POMOST Programme co-financed by the European Union within European Regional Development Fund.
\end{acknowledgments}

\end{document}